\documentclass[useAMS,usenatbib]{mn2e}
\usepackage{graphicx}

\title[Accretion in a Lithium-Rich F Dwarf]{Beryllium Enhancement as
  Evidence for Accretion in a Lithium-Rich F Dwarf}

\author[J. F. Ashwell et al.]{Johanna F. Ashwell,$^{1}$\thanks{E-mail:
jfa@astro.keele.ac.uk} R. D. Jeffries,$^{1}$ Barry Smalley,$^{1}$ C. P. Deliyannis,$^{2}$ \newauthor Aaron Steinhauer$^{2}$ and J. R. King$^{3}$\\
$^{1}$Astrophysics Group, School of Chemistry and Physics, Lennard-Jones Labs, Keele University, Keele, Staffordshire,
ST5~5BG, UK\\
$^{2}$Department of Astronomy, Indiana University, 727 East Third Street,
Bloomington, IN 47405-7105, USA\\
$^{3}$Department of Physics and Astronomy, Clemson University, Kinard Laboratory of Physics, Clemson, SC 29634-0978, USA}

\begin{document}

\date{Accepted 2005 ?? ??. Received 2005 ?? ??; in original form 2005 July 11}

\pagerange{\pageref{firstpage}--\pageref{lastpage}} \pubyear{2005}

\maketitle

\label{firstpage}

\begin{abstract}
The early F dwarf star ``J37'' in the open cluster NGC6633 shows an
unusual pattern of photospheric abundances, including an order of magnitude
enhancement of lithium and iron-peak elements, but an under-abundance
of carbon. As a consequence of its thin convection zone these anomalies
have been attributed to either radiative diffusion or the accretion of
hydrogen-depleted material. By comparing high resolution VLT/UVES
spectra of J37 (and other F stars in NGC\,6633) with syntheses of the
Be {\sc{ii}} doublet region at 3131 \AA, we establish that J37 also has
a Be abundance ($A$(Be)$=3.0\pm0.5$) that is at least ten times the
cosmic value. This contradicts radiative diffusion models that produce
a Li over-abundance, as they also predict photospheric Be
depletion. Instead, since Be is a highly refractory element, it
supports the notion that J37 is the first clear example of a star that
has accreted volatile-depleted material with a composition similar to
chondritic meteorites, although some diffusion may be necessary to
explain the low C and O abundances.

\end{abstract}

\begin{keywords}
Accretion, Diffusion, Stars: Abundances,
Stars: Atmospheres, Stars: Chemically Peculiar, Stars: Evolution.
\end{keywords}

\section{Introduction}
The star J37 [$=$ TYC 445- 1972-1, GSC 00445-01972] is a remarkable
early F-type dwarf member of the Hyades-age open cluster NGC\,6633
(Jeffries 1997). It demonstrates several chemical peculiarities,
including factors of $\simeq 5-10$ enhancement in Fe, Ni and Li and an
under-abundance of carbon (Deliyannis, Steinhauer \& Jeffries 2002 --
hereafter DSJ02; Laws \& Gonzalez 2003 -- hereafter LG03).

DSJ02 suggested that radiatively driven diffusion, specifically the
models of Richer \& Michaud (1993), produce a ``Li-peak'' over a narrow
range of effective temperatures ($6900<T_{\rm eff}<7100$\,K at
800\,Myr), caused by the very thin surface convection zone (SCZ) of J37
reaching down to a radiatively supported reservoir of Li atoms. This
hypothesis explains some (but not all) of the other abundance
anomalies, but is contradicted by recent ``turbulent'' diffusion models
(e.g. Richer, Michaud \& Turcotte 2000) where instead a modest
under-abundance of Li is predicted.

Alternatively, LG03 suggest that accretion of material depleted of
elements with low-condensation temperatures into the thin SCZ can explain
the high abundances of more refractory elements like Li and Fe, whilst
simultaneously resulting in little enrichment of volatiles like C and
O.  A problem here is that C and O may be under-abundant in J37,
perhaps indicating a later role for diffusion processes similar to
those in Am stars.

The Be abundance of J37 would provide an interesting test of these
competing hypotheses. Be is the next lightest metal after Li, but may
behave quite differently in diffusion models due to its differing
electronic structure. The \shortcite{Richer93} diffusion models predict
that at the age and $T_{\rm eff}$ of J37, an over-abundance of Li
should be accompanied by an absence of photospheric Be. In contrast,
because Be has a very high condensation temperature, the accretion of
volatile-depleted material would lead to a Be enhancement as high or
even a little higher than that for Li and Fe.

This letter presents an analysis of high resolution, high
signal-to-noise spectra of J37 and three comparison stars in
NGC\,6633. We derive Be abundances by comparing the spectra with a
model synthesis in the region of the Be\,{\sc ii} doublet at 3130.4\AA\
and 3131.1\AA. We also (re)determine the abundances of several other
elements and compare these to the predictions of diffusion and
accretion models.

\section{Observations}
\begin{table}
  \centering
   \caption{Stellar Parameters for J1, J3, J16 and J37.}
   \label{tab:parameters}
   \begin{tabular}{@{}lcccc}
   \hline
  Star & T$_{eff}$ & log $g$  & $\xi$ & $v$ sin $i$  \\
       & (K)       &(cm\,s$^{-2}$) &  (km\,s$^{-1}$) & (km\,s$^{-1}$)\\

  \hline

  J1   & 6870$\pm$80 & 4.05$\pm$0.15 & 2.29$\pm$0.12 & 19.0$\pm$0.9  \\
  J3   & 6310$\pm$60 & 4.00$\pm$0.20 & 1.53$\pm$0.11 & 10.8$\pm$0.6  \\
  J16  & 6320$\pm$60 & 4.25$\pm$0.15 & 1.27$\pm$0.10 &  7.2$\pm$0.4  \\
  J37  & 7580$\pm$80 & 4.05$\pm$0.15 & 3.33$\pm$0.14 & 29.4$\pm$1.2  \\
 \hline
 \end{tabular}
 \end{table}

High resolution spectra of the NGC\,6633 stars J1, J3, J16 and J37 (see
Jeffries 1997) were obtained with the UV-Visual Echelle Spectrograph
(UVES) mounted on the UT2 8.2-m ESO Very Large Telescope between 13
June 2003 and 06 July 2003. The standard DIC1/346/580 configuration
with a 1.2 arcsec slit was used to take 4 exposures of between 900-s
and 1500-s for each star.  The wavelength range 3050-6820\AA\ (orders
152-90) was covered at a resolving power of 35000. The final
signal-to-noise ratio, achieved after co-adding the exposures,
was 50-70 per 0.03\AA\ pixel in the Be\,{\sc ii} region (at 3131\AA), but much
higher (200-300) at redder wavelengths.

\section{Analysis}
\subsection{Atmospheric Parameters}
\label{sec:parameters}
Our analysis used the Kurucz ATLAS9 model atmosphere grids with no
convective overshoot (Kurucz 1993; Sbordone et al 2004). Target spectra
were examined for suitably unblended Fe {\sc{i}} and Fe {\sc{ii}} lines
(in the red part of the spectra).  The equivalent widths (EWs) of these
were measured in the targets and also in the solar spectrum using the
NOAO solar atlas (Kurucz et al. 1984).  Oscillator strengths were
set assuming the solar parameters T$_{eff}$ = 5777
log $g$ = 4.44, microturbulence $\xi$ = 1.25\,km\,s$^{-1}$, average
metallicity [M/H] = 0.0 and absolute Fe abundance $A$(Fe)$ = 12+ \log
(N$(Fe)/$N($H$)) = 7.54$.  Differential Fe abundances were determined from
these lines in the NGC\,6633 stars along a locus of $T_{\rm eff}$ and
$\log g$ values that gave equal mean abundances from the Fe\,{\sc i}
and Fe\,{\sc ii} lines. Microturbulence values were estimated by
requiring no trend in abundance from the Fe\,{\sc i} lines with EW.

The final atmospheric parameters were settled on by estimating $T_{\rm
eff}$ from a mean of the values given by (i) relationships between
$V-I$ and $V-K$ colour indices and $T_{\rm eff}$ (Alonso, Arribas \&
Mart\'inez-Rodger 1996) and (ii) The combination of $VIJHK$ photometry
and a semi-empirical infra-red flux method using the ATLAS9 models. The
$VI$ photometry comes from Jeffries (1997), $JHK$ photometry comes from
2MASS (Cutri et al. 2003). An average atmospheric metallicity, [M/H],
of 0.0 was assumed for J1, J3 and J16, whilst [M/H]$=+0.5$ was assumed
for J37. In principle we could have iterated these values in the light
of our final abundance estimates, but the reason we chose to use the
$V-I$ and $V-K$ colours is that they are almost insensitive to
metallicity variations in any case.  Extinction and reddening were
accounted for using $E(B-V)=0.165$ from Jeffries et al. (2002) and the
reddening law of Rieke \& Lebofsky (1985). Defining $T_{\rm eff}$ leads
to a $\log g$ from the ionisation balance locus.

Uncertainties in $T_{\rm eff}$ are calculated from the standard error
in the mean of the three estimation methods. These consequently lead through to
additional uncertainties in $\log g$, microturbulence and abundances.
The final atmospheric parameters and an estimate of $v \sin i$
for each star obtained by fitting several of the unblended Fe lines are
given in Table~\ref{tab:parameters}. [Fe/H] values are given in
Table~\ref{tab:abundances}.

\subsection{Blending elements}
The Be\,{\sc ii} doublet region contains many atomic lines.  Because of
finite resolution and rotational broadening, estimates are required for
the abundances of elements that contribute significantly to absorption
in this region -- namely, Ti, V, Mn and Co.  We also included Ca in our
analysis, which is important for distinguishing between accretion and
diffusion scenarios and for which only 1 and 2 lines were analysed by
DSJ02 and LG03 respectively.  The EWs of unblended lines in the red
region of the spectra were measured for these elements, converted into
abundances using the atmospheric parameters described in section
\ref{sec:parameters} and averaged. The results are given in Table
\ref{tab:abundances} along with the number of lines used for each
element. These abundances are differential with respect to the Sun,
because the same procedure for adjusting the oscillator strengths was
performed.  No unblended lines were available for V and Co in
J37. Here, the assumption was made that [V/H] and [Co/H] were equal to
[Fe/H]. We confirmed that assuming a mean cluster abundance for these
elements in J37 would have made little difference to our final derived
Be abundance. Errors in the Ca, Ti and Fe abundances are dominated by
$T_{\rm eff}$ uncertainties.  For Mn, V and Co, the scatter in the
estimates for individual lines is equally important.

The Be\,{\sc ii} region also contains OH and CO molecular lines.  The C
abundance was measured in the same manner as above for J1, J3 and J16
and from a synthesis of the C lines at 5380\AA\ and 6587\AA\ for J37,
using an updated version of the LTE {\sc moog} analysis code (Sneden
2002). The abundance of O in J1 and J37 was estimated using the 7771 -
7775\AA\ O\,{\sc i} triplet EW measurements of LG03, but using our atmospheric
parameters and assuming $A$(O)$= 8.94$.  The quoted differential O
abundances contain non-LTE corrections of -0.51 for J1 and -0.62 for
J37, interpolated from tables in \cite{Takeda97}.

\begin{table*}
  \centering
   \caption{Elemental Abundances in J1, J3, J16 and J37. The number of
   lines/features used in the analyses are indicated in brackets.}
   \label{tab:abundances}
   \begin{tabular}{@{}ll@{\hspace*{1mm}}cl@{\hspace*{1mm}}cl@{\hspace*{1mm}}cl@{\hspace*{1mm}}c}
   \hline
  X       & \ \ [X/H]$_{J1}$&$n$      & \ \ [X/H]$_{J3}$ &$n$       & \
  \  [X/H]$_{J16}$ &$n$     & \ \ [X/H]$_{J37}$ &$n$\\
  \hline
  C       & $-$0.15$\pm$0.03 &(7)  & $-$0.14$\pm$0.03 &(4)  &
  $-$0.11$\pm$0.11 &(5) & $-$0.50$\pm$0.15 &(2)\\
  O$^{a}$ & $+$0.14$\pm$0.05 &(3)  &                  &     &&
  & $-$0.08$\pm$0.15 &(3)\\
  Ca      & $-$0.07$\pm$0.05 &(14)  & $-$0.06$\pm$0.02 &(16)  &
  $-$0.04$\pm$0.03 &(16)  & $+$0.66$\pm$0.04 &(10)\\
  Ti      & $-$0.15$\pm$0.04 &(10)  & $-$0.12$\pm$0.04 &(23)  &
  $-$0.13$\pm$0.04 &(28)  & $+$0.77$\pm$0.10 &(3)\\
  V       & $-$0.18$\pm$0.04 &(3)   & $-$0.26$\pm$0.05 &(4)  &
  $-$0.17$\pm$0.05 &(7) & $+$0.67$^{b}$\\
  Mn      & $-$0.22$\pm$0.08 &(4)   & $-$0.30$\pm$0.06 &(6)  &
  $-$0.34$\pm$0.05 &(6)  & $+$0.54$\pm$0.11 &(3)\\
  Fe      & $-$0.31$\pm$0.04 &(48)  & $-$0.16$\pm$0.03 &(48)  &
  $-$0.15$\pm$0.03 &(58)  & $+$0.67$\pm$0.04 &(40)\\
  Co      & $-$0.10$\pm$0.18 &(3)   & $-$0.17$\pm$0.06 &(5)  &
  $-$0.33$\pm$0.07 &(4)  & $+$0.67$^{b}$\\
  \hline
\multicolumn{9}{l}{$^{a}$ NLTE - Derived from EWs reported by
  LG03.}\\
\multicolumn{9}{l}{$^{b}$ Assumed to scale with [Fe/H] -- no unblended
  lines accessible in the spectrum.}\\
 \end{tabular}
 \end{table*}

\subsection{Be and Li abundances}
\begin{figure}
 \centering
 \includegraphics[width=7.5cm]{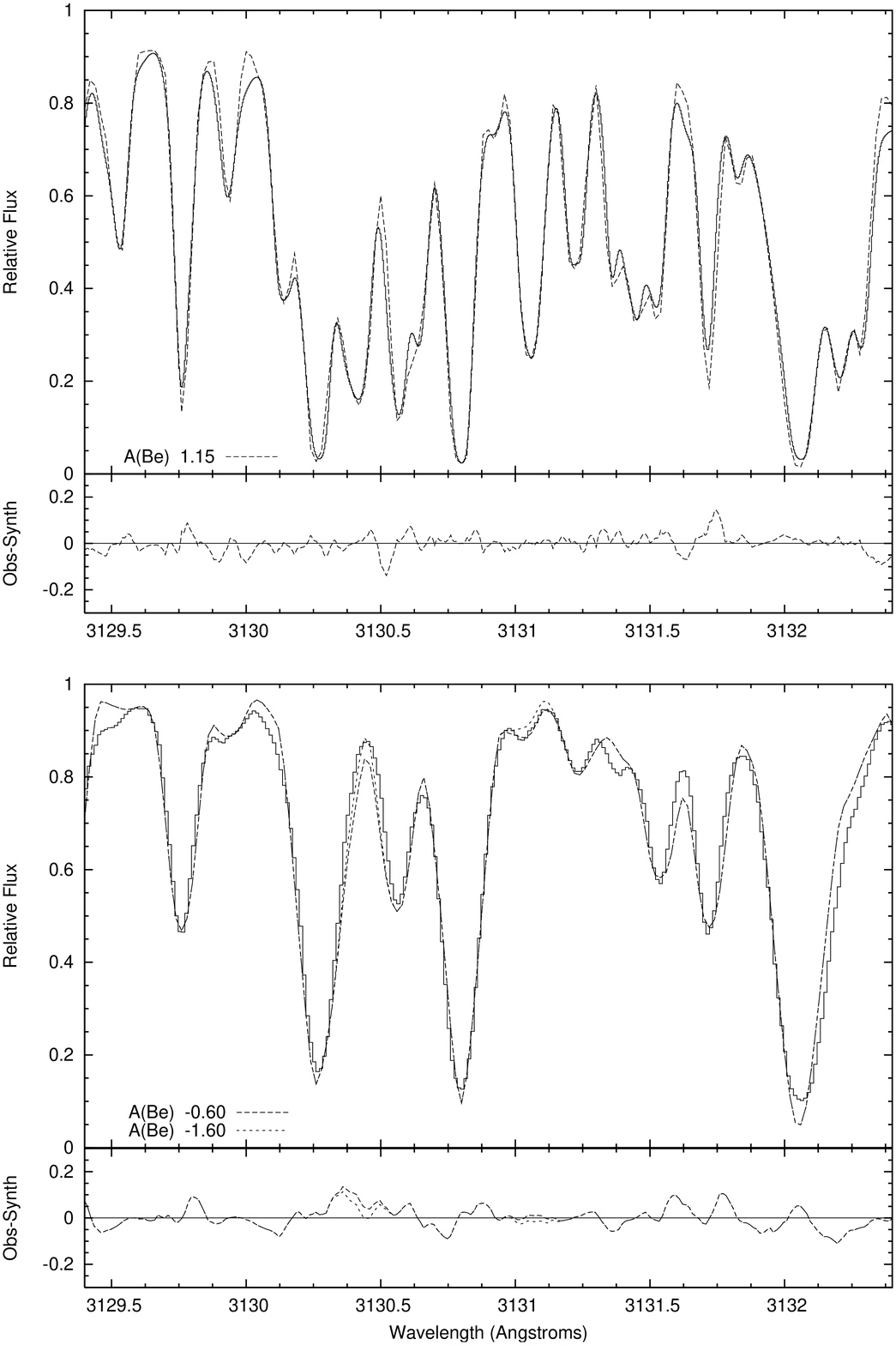}
\caption{Syntheses of the Be {\sc{ii}} doublet (3130.4
\AA \ and 3131.1 \AA) region compared with the NOAO solar atlas and Procyon. The
upper portion of each plot shows the spectrum (solid line) and
synthesis (dashed lines). The lower portions show the
residuals.}
\label{fig:Solar&Procyon}
\end{figure}

LTE Be and Li abundances were determined via spectrum synthesis using
{\sc moog} and atmospheres interpolated from the Kurucz ATLAS9 grids.
The well-established line list in the region of the Li\,{\sc i}
6708\AA\ doublet was taken from Ford et al. (2002).  Atomic and
molecular identifications, wavelengths, excitation potentials and log
$gf$-values used for the Be\,{\sc ii} region were taken from King,
Deliyannis \& Boesgaard (1997), supplemented by molecular lines from
the Kurucz CD-ROMs. Oscillator strengths were tuned to better fit both
the NOAO Solar Atlas {\it and} a spectrum of Procyon (F5\,IV) obtained
with UVES by \shortcite{Bagnulo03} (see Fig.
\ref{fig:Solar&Procyon}).

\begin{figure*}
 \centering
 \includegraphics[angle=-90,width=16.5cm]{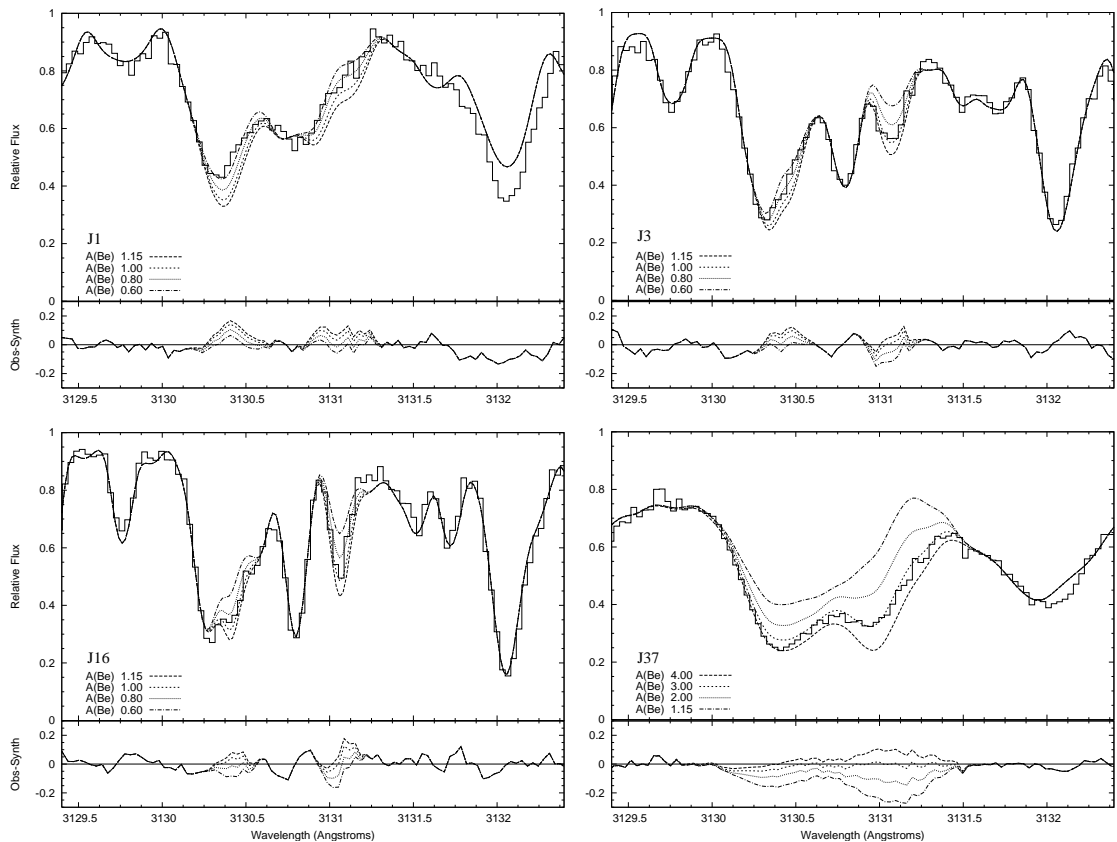}
\caption{Spectral synthesis of the Be {\sc{ii}} doublet region in stars
J1, J3, J16 and J37. In each case the upper plot shows the spectrum
(stepped line) and syntheses with various Be abundances and the lower plot
shows the residuals.}
\label{fig:J_Group}
\end{figure*}

We have largely adopted the corrections discussed in King et al. (1997)
and also the philosophy that log $gf$-values needing the least
adjustment were varied where a number of lines affected the same
feature. The resulting solar $A($Be$) = 1.15$, agreeing with previous
photospheric determinations (Anders \& Grevesse 1989), whilst Procyon
is confirmed as very Be-depleted ($A($Be$)<-0.6$). To simultaneously
match the Procyon spectrum (with assumed T$_{eff}$ = 6700, log $g$ =
4.05) and the Sun, a CH line at 3131.058\AA \ is removed, the
wavelength of the Mn\,{\sc{i}} line at 3131.037\AA \ is shifted to
3131.017\AA \ as suggested by \shortcite{King97} and its log $gf$-value
increased by 1.563 dex. This modified line list provides good fits
(especially in the vicinity of the Be\,{\sc ii} lines) for all the stars in this
study (see Fig.~\ref{fig:J_Group}).

The estimated LTE Be and Li abundances are listed in Table
\ref{tab:LiBe_abundances}. The enormous Li over-abundance (compared with
a meteoritic value of $A($Li$)=3.3$) of J37 is confirmed. The Li
abundances of J1, J3 and J16 are consistent with an initial meteoritic
abundance which has undergone some depletion as a result of internal
mixing of Li. The Be abundances of J1, J3 and J16 also appear depleted
with respect to both solar ($A($Be$)=1.15$) and meteoritic
($A($Be$)=1.42$) abundances. This has been seen in other Hyades-age
clusters (e.g. Boesgaard, Armengaud \& King 2003), although the large
ratio of Li to Be in J1 and J3 suggests either that more Be than Li has
been depleted, which seems unlikely, or that the initial Li to Be ratio in NGC\,6633 may be a
few tenths of a dex higher than in meteorites or other young
clusters. However, this uncertainty in the initial Be content of
NGC\,6633 is dwarfed by the massive over-abundance of Be estimated for
J37, where the quoted error is dominated by fitting uncertainties
rather than atmospheric parameter uncertainties.

\begin{table}
  \centering
   \caption{Abundances of Li and Be for J1, J3, J16 and J37.}
   \label{tab:LiBe_abundances}
   \begin{tabular}{@{}lcccc}
   \hline
  X & A(X)$_{J1}$ & A(X)$_{J3}$ & A(X)$_{J16}$ & A(X)$_{J37}$ \\
  \hline
  Li &  2.90$\pm$0.11 &  3.10$\pm$0.14 &  2.70$\pm$0.14 &  4.40$\pm$0.14 \\
  Be &  0.70$\pm$0.15 &  0.90$\pm$0.15 &  0.90$\pm$0.15 &  3.00$\pm$0.50 \\
  \hline
 \end{tabular}
\end{table}

\section{Discussion}

This letter improves on the work of DSJ02 and LG03 in several respects. The
stellar parameter determinations are more secure thanks to (i) the use
of metallicity insensitive colour indices for the $T_{\rm eff}$
determination; (ii) 34-51 Fe {\sc{i}} lines for the metallicity and
microturbulence estimates; (iii) a total of 40-58 Fe lines for the $\log g$
determination from the ionisation balance. The determination of
abundance anomalies for J37 with respect to the cluster are improved
by the use of 3 cluster comparison stars (LG03 used only J1); the
addition of Be and Mn abundances; a more reliable determination of the
C and Ca abundances using more spectral features; and an apparent
mistake in the application of NLTE O-abundance corrections by LG03 has
been rectified.

Assuming an initial $A$(Be)$=1.2\pm0.2$ for NGC\,6633, then J37 is
enhanced in Be by $1.8\pm 0.5$ dex (according to our LTE
analysis). Likewise, we confirm previous results that Li is enhanced by
about $1.10 \pm 0.25$\,dex (assuming an initial $A$(Li)$=3.3\pm0.2$),
Fe by about $0.85\pm0.06$\,dex (compared with the average [Fe/H] of the
three comparison stars), Ti by $0.90\pm0.10$\,dex, Mn by
$0.83\pm0.12$\,dex, and Ca by $0.72\pm0.04$\,dex.  These
over-abundances are larger than found by DSJ02 and LG03, because the
$T_{\rm eff}$ we have derived is about 500\,K higher than indicated by
the $B-V$ based estimate in DSJ02 and 275\,K hotter than the
spectroscopic estimate in LG03 (about 1 error bar). Given the abundance
anomalies in J37's atmosphere it would not be surprising if a $T_{\rm
eff}$ based on $B-V$, which is heavily affected by line blanketing, was
inaccurate.  The optical/IR indices we have used are much more
insensitive to detailed composition.

The large Be abundance found for J37 offers no support to the diffusion
hypothesis put forward by DSJ02. The only set of diffusion models that
predicted a large Li enhancement also predicted that Be should be
completely depleted from the photosphere (Richer \& Michaud
1993). Furthermore, the new $T_{\rm eff}$ we have estimated for J37
places it $>400$\,K hotter than the predicted position of the Li-peak
at the age of NGC\,6633. The extreme sensitivity of diffusion predictions to details of the
atmosphere and atomic physics may mean that future calculations are
more successful; but currently published models fail.

The accretion hypothesis can satisfactorily explain why there are
simultaneous increases of Be, Li and Fe but not of C and O. Be has a
higher condensation temperature than either Fe or Li (Lodders 2003) and
should be present at close to its primordial value in any
volatile-depleted material that is accreted.

Taking Fe as the primary compositional tracer and making the
simplifying assumptions that accreted material is mixed only within the
SCZ and no diffusion or gravitational settling takes place after
accretion, we can estimate the mass of accreted material that would
explain the observed abundance. According to the models of Richer et
al.  (2000), the $T_{\rm eff}$ of J37 matches that for a star of
1.6\,$M_{\odot}$ at an age of 600\,Myr and such a star would have $\log
g =4.15$, in excellent agreement with our spectroscopically determined
value. The mass of the SCZ in such a star is of order
$10^{-6}\,M_{\odot}$ and so the quantity of (for instance) solar system
CI chondrite meteoritic material (Lodders
2003) required for the Fe enhancement observed in J37 (assuming an
initial A(Fe) = 7.39$\pm$0.06 determined from the
comparison stars) would be just 0.011\,$M_{\oplus}$.

The accretion of this chondritic material would also result in
$A$(Li)$=4.04$ and $A$(Be)$=2.13$ (assuming the initial abundances
quoted earlier).  In both cases the uncertainty in the initial
abundance has little effect on the final abundance and hence it appears
that the measured Li/Fe and Be/Fe ratios in J37 might be too large to
support the accretion hypothesis (see
Table~\ref{tab:LiBe_abundances}). Recall however that these abundances
are calculated assuming LTE conditions.  The Li\,{\sc i}~6708\AA\
doublet is strongly effected by NLTE effects at high abundances as can
be seen in Carlsson et al. (1994, see their figure 16). They suggest an
NLTE correction to the J37 Li abundance of $-0.3$ dex bringing the Li
abundance in line with that predicted by meteoritic accretion. The Be
abundance may also be effected by atmospheric uncertainties.  Detailed
NLTE calculations at a range of $T_{\rm eff}$ do not yet exist and
there is also a strong suggestion (e.g. see Asplund 2004) that there is
a missing source of continuum opacity at $\simeq 3130$\AA. Given that
J37 is significantly hotter than the Sun and the NGC\,6633 comparison
stars, these two effects result in additional uncertainty in the
differential Be abundances, to the extent that detailed agreement with
the meteoritic accretion hypothesis should not be excluded.

Accretion of volatile-depleted meteoritic material cannot explain why
the C and O abundances of J37 are {\em lower} than the NGC\,6633
comparison stars by 0.2-0.3\,dex.  The accretion of 0.011\,$M_{\oplus}$
of chondritic material should result in [C/H] = $+0.02\pm0.05$ and
[O/H] = $+0.44\pm0.05$ (assuming initial abundances of $-$0.13 and
$+$0.14 respectively), about 0.5\,dex higher than observed in J37. This
discrepancy can only be partly resolved by accreting material that is
even more depleted in volatiles (such as ``bulk Earth'' material -- see
LG03). It seems likely, as suggested by LG03, that some diffusion {\em
has} occurred in J37, either before or after the accretion
event(s). Certainly C and O depletions of order 0.5\,dex are feasible
according to the recent models of Richer et al. (2000) and are less
severe than the depletions observed in many classic Am/Fm stars.
However, given that models and also observations of Am stars, show
simultaneous depletions of similar magnitude for the refractory element
Ca, accompanied by an enhancement of the iron-peak elements, it seems
unlikely that diffusion can have been very effective, because
[Ca/Fe]\,$\simeq -0.1$ for J37, rather than the more typical $-1$ seen in
Am stars.

An area for further study would be to systematically find the frequency
of Li/Be-rich F stars in open clusters, to investigate (a) how common
the accretion events are and (b) how quickly diffusive or other mixing
processes can remove the abundance anomalies from the photosphere.  If
the frequency is low it would suggest that accretion events are rare or
that the effects of diffusion rapidly dilute any accreted material. If,
however, the frequency is high it would suggest accretion is common and
diffusion is slow.  It should be noted that the quantities of accreted
material discussed in this paper are far too small to explain the
enhanced metallicity of exoplanet host stars (e.g. Santos, Israelian \&
Mayor 2004), which generally have {\it much} more massive SCZs.

\section{Summary}
We have found that that the Li-rich early F dwarf J37 is also Be-rich,
with $A$(Be)$=3.0\pm0.5$. The Be over-abundance in J37 (together with
0.7--0.9\,dex enhanced abundances for Ca and iron-peak elements)
completely contradict published diffusion models hypothesised to explain the
Li-rich nature of J37. Instead, the enhancement of highly refractory
elements like Be, Ca and Fe, together with much lower abundances of C
and O, can be explained by the accretion of about 0.01\,$M_{\oplus}$ of
volatile-depleted material such as chondritic meteorites. However, the
$\sim 0.3$\,dex under-abundances of C and O in J37 compared with other
stars in the same cluster suggest that there may still be a limited
role for diffusion to play.

\section*{Acknowledgements}
Based on observations collected with the VLT/UT2 Kueyen telescope
(Paranal Observatory, ESO, Chile, observing runs 071.D-0441 and
266.D-5655). JFA acknowledges the financial support of the UK Particle
Physics and Astronomy Research Council (PPARC). This research has made
use of the SIMBAD Database, operated at CDS, Strasbourg, France and
NASA's Astrophysics Data System.

\end{document}